# Modeling and Analysis of Wireless Channels via the Mixture of Gaussian Distribution

Bassant Selim, *Student Member, IEEE,* Omar Alhussein, *Student Member, IEEE,*

Sami Muhaidat, *Senior Member, IEEE,* George K. Karagiannidis, *Fellow, IEEE*

and Jie Liang, *Senior Member, IEEE*

### Abstract

In this paper, we consider a unified approach to model wireless channels by a mixture of Gaussian (MoG) distributions. The proposed approach provides an accurate approximation for the envelope and the signal-to-noise ratio (SNR) distributions of wireless channels. Simulation results have shown that our model can accurately characterize multipath fading and composite fading channels. We utilize the well known expectation-maximization algorithm to estimate the parameters of the MoG distribution and further utilize the Bayesian information criterion to determine the number of mixture components automatically. We employ the Kullback-Leibler divergence and the mean square error criteria to demonstrate that our distribution provides both high accuracy and low computational complexity. Additionally, we provide closed-form expressions or approximations for several performance metrics used in wireless communication systems, including the moment generating function, the raw moments, the amount of fading, the outage probability, the average channel capacity, and the probability of energy detection for cognitive radio. Numerical Analysis and Monte-Carlo simulation results are presented to corroborate the analytical results.

Bassant Selim and George K. Karagiannidis are with the Department of Electrical and Computer Engineering, Khalifa University, PO Box 127788, Abu Dhabi, UAE, e-mails: bassant.selim, george.karagiannidis@kustar.ac.ae.

Omar Alhussein, Sami Muhaidat, and Jie Liang are with the School of Engineering Science, Simon Fraser University, Burnaby, BC, Canada, V5A1S6, e-mails: oalhusse@sfu.ca, muhaidat@ieee.org, jiel@sfu.ca.

George K. Karagiannidis is with the Department of Electrical and Computer Engineering, Aristotle University of Thessaloniki, 54124, Thessaloniki, Greece, e-mail: geokarag@ieee.org.

# I. INTRODUCTION

Modeling the terrestrial wireless propagation is of importance for the design and performance analysis of wireless systems. In a typical mobile radio propagation scenario, the received signal presents small scale power fluctuations, due to multipath propagation, superimposed on large scale signal power fluctuations, also known as shadowing, which is due to the presence of large obstacles between the transmitter and receiver. The small scale fading results in very rapid fluctuations around the mean signal level, while shadowing gives rise to relatively slow variations of the mean signal level [1]. A common example of composite fading channels is the Nakagami-Lognormal (NL) channel. In this case, the density function is obtained by averaging the instantaneous Nakagami-$m$ fading average power over the conditional probability density function (pdf) of the log-normal shadowing, resulting in a complicated pdf that has no closed form expression [2].

The $K$ [3] and generalized-$K$ ($K_G$) distributions [4], [5], have been introduced as relatively simpler models to characterize composite fading channels, in which the Lognormal distribution is replaced by the Gamma distribution in the Rayleigh-Lognormal (RL) and NL distributions, respectively. The $\kappa - \mu$ and the $\eta - \mu$ distributions [6] are general fading distributions for line-of-sight (LOS) and non-line-of-sight applications, respectively. These distributions can represent the Rice (Nakagami-$n$), the Nakagami-$m$, the Rayleigh, the One-Sided Gaussian, and the Hoyt (Nakagami-$q$) distributions as special cases. Quite recently, a generalization of the $\kappa$-$\mu$ fading channel, where the LOS component is shadowed, has been proposed in [7]. All these models contain the modified Bessel function of the first or second kind, which complicates further analytical performance measures. In [8], the Lognormal distribution was replaced by the Inverse-Gaussian distribution, resulting in the Rayleigh/Inverse Gaussian (RIGD) distribution, followed by its generalized versions, i.e., the $\mathcal{G}$-distribution [9], the $\kappa - \mu$/Inverse Gaussian distribution [10] and the $\eta - \mu$/Inverse Gaussian distribution [11]. The drawback of these distributions is their increased complexity due to the presence of the modified Bessel function of the second kind. Recently, an interesting work has been proposed by Atapattu *et al*. [12], where several channel models were expressed as a mixture Gamma ($M\mathcal{G}$) distribution via Gauss-Quadrature approximation. The $M\mathcal{G}$ model is more accurate than the aforementioned alternatives, and it has the advantage of simplicity as well.

Finite mixtures of distributions provide a mathematical-based approach to statistical modeling of a wide variety of random phenomena [13]. In this paper, an alternative model, that represents both composite and non-composite fading channels by Mixture of Gaussian (MoG) distributions is presented. The approximation method is based on the expectation-maximization (EM) algorithm, which was coined by Dempster *et al.* in their seminal paper [14]. The EM algorithm is essentially a set of algorithms exceptionally useful for finding the maximum likelihood estimator (MLE) of any distribution in the exponential family [15], and widely used for the missing data problem (i.e., modeling a mixture distribution). The main contributions of this paper can be summarized as:

- We propose MoG distributions to model both the envelope and the signal-to-noise ratio (SNR) distributions of wireless channels. The proposed approach is proven to accurately model both composite and non-composite channels in a very simple expression.

- We determine the number of components using the Bayesian information criterion (BIC), while the corresponding parameters for the mixture are evaluated using the EM algorithm.

- We demonstrate the importance and tractability of our model by deriving several tools for the performance analysis of single-user communications such as the outage probability and raw moments. Moreover, we derive the moment generating function (MGF), of which the symbol error rate (SER) of $L$-branch maximal ratio combining (MRC) diversity system is presented for various signaling schemes. Finally, we derive an approximation for the average detection probability in cognitive radio networks and provide an upper bound to the error.

- Numerical analysis and Monte Carlo simulation results are presented to corroborate the derived analytical results.

The rest of this paper is organized as follows. Section II gives a brief description of several wireless channel models of interest. In Section III, the MoG distribution is introduced together with a brief description of the EM algorithm. Section IV presents a detailed comparison of the MoG distribution to the channel models it can approximate. In Section V, performance metrics, such as the MGF, the raw moments, the amount of fading, the outage probability, and the average channel capacity are derived using the MoG distribution. Simulation results and numerical analysis are presented in Section VI. Finally Section VII concludes this work.

## II. FADING CHANNELS

Radio-wave propagation through wireless channels undergoes detrimental effects characterized by multipath fading and shadowing. Modeling of such fading channels is typically a complex process and often leads to intractable solutions. Considerable efforts have focused on the statistical modeling which resulted in a wide range of statistical models for fading channels [2].

In this section, we give a brief description of some well known channels that often lead to intractable performance analysis of wireless communication systems.

### A. The Nakagami-Lognormal Channel

The NL fading model is a mixture of Nakagami-$m$ distribution and Lognormal distribution obtained by averaging the instantaneous Nakagami-$m$ fading average power over the pdf of the log-normal shadowing as follows

$$f_\alpha\left(\alpha\right) = \int_0^\infty f_\alpha\left(\alpha|\sigma\right) f_\sigma\left(\sigma\right) \, d\sigma, \tag{1}$$

where $f_\alpha(\alpha|\sigma)$ is the Nakagami-*m* distribution given by

$$f_\alpha\left(\alpha|\sigma\right) = \frac{2m^m}{\sigma^m \Gamma\left(m\right)} \alpha^{2m-1} e^{-m\frac{\alpha^2}{\sigma}}, \tag{2}$$

where $\Gamma(.)$ is the gamma function [16] and $m$ is the fading parameter, which is inversely proportional to multipath fading severity i.e., as $m \to \infty$, multipath severity diminishes. Note here that $m = \frac{\mathbb{E}^2\{\gamma\}}{\text{Var}\{\gamma\}}$. The average power $\sigma$ follows a Lognormal distribution, contributing to shadowing at longer routes, expressed as

$$f_\sigma\left(\sigma\right) = \frac{e^{\frac{-(10\log(\sigma)-M)^2}{2\zeta^2}}}{\sqrt{2\pi}\sigma\lambda\zeta}, \tag{3}$$

where $\lambda = \frac{\ln 10}{10}$, $M$ and $\zeta^2$, measured in dB, are the mean and variance of the Gaussian RV $V = 10\log_{10}(\sigma)$, respectively. In order to compare (2) with that of the Gaussian RV $X = \ln(\sigma)$, the following relations apply [17]

$$X = \lambda V, \tag{4}$$

$$M_X = \lambda M,$$

$$\zeta_X = \lambda\zeta.$$

An important remark regarding the Lognormal distribution is that, while $\zeta$ essentially defines different Lognormal distributions, $M$ is effectively a scaling factor [17]. Denoting $M_n = 10^{\frac{M}{10}}$ and $x = \frac{\sigma}{M_n}$, then it is straightforward to show that

$$f_\alpha\left(\alpha M_n\right) = \frac{1}{M_n} f_\alpha\left(\alpha|M = 0\right). \tag{5}$$

Therefore, it is only sufficient to perform an approximation for $M = 0$ dB, and generalize the results to other scaling factors.

Let $E_s$ denote the energy per symbol, $N_0$ be the single sided power spectral density of the complex additive white Gaussian noise (AWGN) and assuming $\mathbb{E}\left[|\alpha^2|\right] = 1$, where $\mathbb{E}[.]$ denotes the expectation operator. By applying the following transformation to (1)

$$\gamma = \alpha^2 \overline{\gamma}, \tag{6}$$

 

where $\overline{\gamma} = \mathbb{E}[\gamma] = \frac{E_s}{N_0}$ is the average SNR, we obtain the NL distribution as

$$f_\gamma(x) = \frac{(8.686)\, m^m}{\Gamma(m)\sqrt{2\pi}\zeta} \int_0^\infty \frac{x^{m-1}}{\overline{\gamma}^m \sigma^{m+1}} e^{-\frac{mx}{\overline{\gamma}\sigma}} e^{\frac{-(20\log\sigma)^2}{2\zeta^2}}\, da. \tag{7}$$

The SNR density function is not expressed in a closed form, making the performance analysis of wireless communications under this particular channel very complicated or intractable. Note that the RL distribution is a special case of NL distribution with $m = 1$.

### B. The $\kappa - \mu$ and $\eta - \mu$ Fading Models

With the emergence of relatively new wireless mediums, such as which occurs in underwater acoustic [18], [19] and body communication [20]–[23] fading channels, the wireless research community have had a reincarnated interest in finding more accurate and generalized fading models that provide a better fit to new and realistic measurements. Consequently, new fading models, such as the $\kappa - \mu$ and $\eta - \mu$ distributions, were proposed [6]. The $\kappa - \mu$ fading model is mostly used to represent the multipath fading with LOS condition and includes the following fading models as special cases: the Nakagami-$n$ (Rice), the Nakagami-$m$, the Rayleigh, and the One-sided Gaussian. The instantaneous $\kappa - \mu$ SNR distribution is expressed as [6].

$$f_\gamma(x) = \frac{\mu\left(\frac{1+\kappa}{\overline{\gamma}}\right)^{\frac{\mu+1}{2}}}{\kappa^{\frac{\mu-1}{2}}\exp(\mu\kappa)} x^{\frac{\mu-1}{2}} \exp\left(-\frac{\mu(1+\kappa)}{\overline{\gamma}}x\right) I_{\mu-1}\left(2\mu\sqrt{\frac{\kappa(1+\kappa)x}{\overline{\gamma}}}\right), \tag{8}$$

where $\kappa > 0$ is the ratio between the total power of the dominant components and the total power of the scattered waves, $\mu > 0$ is given by $\mu = \frac{\mathbb{E}^2\{\gamma\}}{\mathrm{Var}\{\gamma\}}\frac{(1+2\kappa)}{(1+\kappa)^2}$, and $I_\mu(.)$ is the modified Bessel function of the first kind and order $\mu$ [16, eq. 8.445]. It is worth mentioning that as $\kappa$ tends to zero, the $\kappa - \mu$ distribution degenerates to the exact Nakagami-$m$ distribution, with $\mu = m = \frac{\mathbb{E}^2\{\gamma\}}{\mathrm{Var}\{\gamma\}}$. Additionally, by setting $\mu = 1$, the $\kappa - \mu$ distribution degenerates to the exact Nakagami-$n$ distribution, with $\kappa = n$.

Complementing the $\kappa - \mu$ model, the $\eta - \mu$ model was proposed to represent NLOS multipath environments, where it includes the Nakagami-$q$, the Nakagami-$m$, and the One-Sided Gaussian distributions as special cases. The instantaneous $\eta - \mu$ SNR distribution is expressed as [6].

$$f_{\gamma}\left(x\right) = \frac{2\sqrt{\pi}\mu^{\mu+\frac{1}{2}}h^{\mu}}{\Gamma\left(\mu\right)H^{\mu-\frac{1}{2}}\overline{\gamma}^{\mu+\frac{1}{2}}}x^{\mu-\frac{1}{2}}\exp\left(-\frac{2\mu h x}{\overline{\gamma}}\right)I_{\mu-\frac{1}{2}}\left(\frac{2\mu H}{\overline{\gamma}}x\right),\qquad(9)$$

where $\mu > 0$ is given by $\mu = \frac{1}{2}\frac{\mathbb{E}^2\{\gamma\}}{\text{Var}\{\gamma\}} + \frac{1}{2}\frac{\mathbb{E}^2\{\gamma\}}{\text{Var}\{\gamma\}}\left(\frac{H}{h}\right)^2$, and parameters $h$ and $H$ can have two different formats corresponding to two different physical phenomena as follows: In Format 1, $h = \frac{2+\eta^{-1}+\eta}{4}$ and $H = \frac{\eta^{-1}-\eta}{4}$, where $0 < \eta < \infty$ is interpreted as the power ratio between the independent in-phase and quadrature components. In Format 2, the in-phase and quadrature components are correlated and have a power ratio of unity. The two corresponding parameters are defined by $h = \frac{1}{1-\eta^2}$ and $H = \frac{\eta}{1-\eta^2}$, where $-1 < \eta < 1$ represents the correlation between the in-phase and quadrature components. The two formats can be obtained from each other using the relation $\eta_{\text{Format1}} = \frac{1-\eta_{\text{Format2}}}{1+\eta_{\text{Format2}}}$. It is worth mentioning that the $\eta - \mu$ distribution degenerates to the Nakagami-$q$ distribution by setting $\mu = 0$, with $q = \sqrt{\eta}$ in Format 1 and $q = \sqrt{\frac{1-\eta}{1+\eta}}$ in Format 2.

### C. The $\kappa - \mu$ Shadowed Fading Models

The $\kappa - \mu$ Shadowed fading model, was firstly proposed as a LOS Shadow fading model [7], where unlike the NL formulation above, it is assumed that only the dominant components of the multipath clusters are subject to random fluctuations. The unconditional instantaneous SNR distribution of the $\kappa - \mu$ Shadowed model is obtained by averaging the conditional $\kappa - \mu$ distribution over the Nakagami-$m$ distribution as follows

$$\begin{aligned}f_{\gamma}\left(\gamma\right) &= \int_0^{\infty} f_{\gamma|\xi}\left(\gamma;\xi\right)f_{\xi}\left(\xi\right)d\xi\\ &= \frac{\mu\left(1+\kappa\right)^{\frac{\mu+1}{2}}}{\overline{\gamma}\kappa^{\frac{\mu-1}{2}}}\left(\frac{\gamma}{\overline{\gamma}}\right)^{\frac{\mu-1}{2}}\exp\left(-\frac{\mu\left(1+\kappa\right)\gamma}{\overline{\gamma}}\right)\frac{m^m}{\Gamma\left(m\right)}\Theta\left(\gamma\right),\end{aligned}\qquad(10)$$

where $\Theta\left(\gamma\right) \triangleq \int_0^{\infty}2\exp\left(-\xi^2(\mu\kappa+m)\right)\xi^{2m-\mu}I_{\mu-1}\left(2\mu\xi\sqrt{\frac{\kappa(1+\kappa)\gamma}{\overline{\gamma}}}\right)d\xi$, which results in the following closed-form expression

$$f_{\gamma}\left(\gamma\right) = \frac{\mu^{\mu}m^m\left(1+\kappa\right)^{\mu}}{\Gamma\left(\mu\right)\overline{\gamma}\left(\mu\kappa+m\right)^m}\left(\frac{\gamma}{\overline{\gamma}}\right)^{\mu-1}\exp\left(-\frac{\mu\left(1+\kappa\right)\gamma}{\overline{\gamma}}\right){}_1F_1\left(m,\mu;\frac{\mu^2\kappa\left(1+\kappa\right)}{\mu\kappa+m}\frac{\gamma}{\overline{\gamma}}\right),\quad(11)$$

  

where the function $_1F_1(.,.;.)$ is the confluent hypergeometric function [16, eq. 9.210.1], defined as $_1F_1(a,b;z) = 1 + \frac{a}{b}\frac{z}{1!} + \frac{a(a+1)}{b(b+1)}\frac{z^2}{2!} + \frac{a(a+1)(a+2)}{b(b+1)(b+2)}\frac{z^3}{3!} + ....$

Interestingly, in a very recent work [24], it has been shown that under a new formulation the $\kappa - \mu$ Shadowed fading model can represent the $\eta - \mu$ distribution as a special case with $\underline{\mu} = 2\mu$, $\underline{\kappa} = \frac{1-\eta}{2\eta}$, and $\underline{m} = m$, where the underlined symbols belong to the $\kappa - \mu$ Shadowed model for the sake of clarity.

## III. The MoG Distribution

We consider the problem of estimating the wireless channels' density functions. Gaussian mixtures [25]–[29] are often used due to the fact that their individual densities are efficiently characterized by the first two moments [30], [31]. The MoG distribution is attributed to have the Universal-approximation property, as it has been proven by Weiners approximation theorem [25], which states that the MoG distribution can approximate any arbitrarily shaped non-Gaussian density. The objective of this section is to provide a unified MoG distribution that can accurately represent different fading channels.

### A. Parameter Estimation of the MoG Distribution

Let the $i^{th}$ entry of a random data vector $Y = (y_1, .., y_n)$, which represents the channel fading amplitude of the composite models, be regarded as incomplete data and modeled as a finite mixture of Gaussians as follows

$$p(y_i|\theta) = \sum_{j=1}^{C} \omega_j \phi(y_i, \theta_j), \ \ y_i \geq 0 \tag{12}$$

where $i = 1, ..., n$ and $C$ represents the number of components. Each $j^{th}$ component is expressed as

$$\phi(y_{i,}\theta_j) = \frac{1}{\sqrt{2\pi}\eta_j}\exp\left(-\frac{(y_i - \mu_j)^2}{2\eta_j^2}\right), \tag{13}$$

where the weight of the $j^{th}$ component is $\omega_j > 0$, with $\sum_j^C \omega_j = 1$. The parameter $\theta_j = \left( \mu_j, \eta_j^2 \right)$ correspond to the mean and variance of the $j^{th}$ component, respectively.

Let the complete data $X$ be the joint probability between $Y$ and $Z$, where $Z \in \{1, .., C\}$ is a hidden (latent) discrete RV that defines which Gaussian component the data vector $Y$ comes from, namely,

$$p\left(Z = j\right) = \omega_j, \; j = 1, .., C. \tag{14}$$

Ideally, one would like to maximize the log-likelihood function as follows

$$\begin{align} \theta_{MLE} &= \arg\max_{\theta \in \Theta} \mathrm{L}\left(\theta\right) \tag{15} \\ &= \arg\max_{\theta \in \Theta} \log\, p\left(y|\theta\right). \end{align}$$

However, maximizing $\mathrm{L}\left(\theta\right)$ is not tractable and difficult to optimize [32]. Instead, the EM algorithm solves the MLE problem by maximizing the so-called $\mathcal{Q}$-function as follows [33]

$$\begin{align} \theta^{(m+1)} &= \arg\max_{\theta \in \Theta} \mathcal{Q}\left(\theta|\theta^{(m)}\right) \tag{16} \\ &= \arg\max_{\theta \in \Theta} \mathbb{E}_{X|y,\theta^{(m)}}\left[\log p_X\left(X|\theta\right)\right], \end{align}$$

where $m$ is the iteration index. The EM algorithm is performed by two iterative steps, namely the expectation step ($\mathbb{E}$-step), and the maximization step ($\mathbb{M}$-step). We set initial guesses of the MoG coefficients, i.e. $\omega^{(0)}, \mu^{(0)}, \eta^{(0)}$, whereby in the $\mathbb{E}$-step, we compute the posterior probability (membership probability)

$$\rho_{ij}^{(m)} = \frac{\omega_j^{(m)}\phi\left(y_i|\mu_j^{(m)},\eta_j^{(m)}\right)}{\sum_{l=1}^C \omega_l^{(m)}\phi\left(y_i|\mu_l^{(m)},\eta_l^{(m)}\right)}. \tag{17}$$

In the $\mathbb{M}$-step, the coefficients are updated by differentiating the $\mathcal{Q}$-function with respect to $\omega$, $\mu$, and $\eta$, resulting in the following analytical $(m+1)^{th}$ estimates

$$\omega_j^{(m+1)} = \frac{1}{n}\sum_{i=1}^n \rho_{ij}^{(m)}, \; j = 1, ..., C, \tag{18}$$

$$\mu_j^{(m+1)} = \frac{1}{n_j^{(m)}}\sum_{i=1}^n \rho_{ij}^{(m)} y_i, \; j = 1, ..., C, \tag{19}$$

 

$$\eta_j^{(m+1)} = \frac{1}{n_j^{(m)}} \sum_{i=1}^{n} \rho_{ij}^{(m)} \left( y_i - \mu_j^{(m+1)} \right)^2, \ j = 1, .., C. \tag{20}$$

This iterative procedure is terminated upon convergence, that is when $\left| L^{(m+1)} - L^{(m)} \right| < \delta$, where

$$L^{(m)} = \frac{1}{n} \sum_{i=1}^{n} \log \left( \sum_{j=1}^{C} \omega_j^{(m)} \phi \left( y_i | \mu_j^{(m)}, \eta_j^{(m)} \right) \right), \ i = 1, ..., n, \tag{21}$$

is the log-likelihood, and $\delta$ is a preset threshold.

The EM algorithm is guaranteed not to get worse as it iterates, i.e. $L^{(m+1)} \leq L^{(m)}$ [14]. Hence, the lower $\delta$ is set, the more accurate the approximation would be. In addition, one can always increase the accuracy by increasing the number of components. Though, this technique might be stuck in a local maxima, since the likelihood is a marginal distribution. However, one could mitigate this problem by heuristics and multiple initial guesses. In this regard, Do *et al.* [32] suggest to initialize parameters in a way that breaks symmetry in mixture models. Finally, it is noteworthy to point out that the EM algorithm has an advantage of being a completely unsupervised learning algorithm, which makes it very convenient for our density estimation application. For more details, one can refer to [15], [33], and references therein.

### B. The pdf of the Instantaneous SNR of the MoG Model

With the aid of the EM algorithm, all fading channels' amplitudes can be represented as

$$f_\alpha (x) = \sum_{j=1}^{C} \frac{\omega_j}{\sqrt{2\pi}\eta_j} \exp \left( -\frac{(x - \mu_j)^2}{2\eta_j^2} \right), \ \ x \geq 0. \tag{22}$$

By taking the change of variables $\gamma = \overline{\gamma} x^2$, the pdf of the instantaneous SNR of the MoG distribution can be written as

$$f_\gamma (\gamma) = \sum_{j=1}^{C} \frac{\omega_j}{\sqrt{8\pi\overline{\gamma}}\eta_j} \frac{1}{\sqrt{\gamma}} \exp \left( \frac{-\left( \sqrt{\frac{\gamma}{\overline{\gamma}}} - \mu_j \right)^2}{2\eta_j^2} \right), \ \ \gamma \geq 0. \tag{23}$$

*C. Determining the optimal number of mixture components*

Generally, when fitting a finite mixture distribution, the determination of an appropriate number of mixture components is inevitably a necessity. Choosing a small number of components would yield an inaccurate representation, while a very large number of components would unnecessarily increase the complexity of the distribution and may cause over-fitting. In addition, Chen [34] has shown that knowledge of the the number of components yields a faster optimal convergence rate for the estimates of a finite mixture than it would when the number of components is unknown.

In this subsection, in order to derive an appropriate number of mixture components, we adopt a simple yet effective unsupervised information theoretic criterion, called the BIC, which was introduced by Gideon Schwarz in [35].

Let $\boldsymbol{x} = \{x_1, ..., x_z, ..x_{\mathcal{M}}\}$, correspond to $\mathcal{M}$ independent and identically distributed (*i.i.d.*) samples, drawn from any of the envelope distributions of the actual aforementioned fading models, then the log-likelihood function of the MoG distribution can be expressed as

$$\mathbb{L}_C \left( \hat{\boldsymbol{\theta}} \right) = \ln \Pr \left( \boldsymbol{x} | \hat{\boldsymbol{\theta}}, C \right) = \sum_{z=1}^{\mathcal{M}} \ln \left\{ \sum_{i=1}^{C} \frac{\hat{\omega}_i}{\sqrt{2\pi}\hat{\eta}_i^2} \exp \left( -\frac{(x_z - \hat{\mu}_i)^2}{2\hat{\eta}^2} \right) \right\} \quad (24)$$

where $\hat{\boldsymbol{\theta}} = [\hat{\omega}_1, \hat{\omega}_2, ..., \hat{\omega}_C, \hat{\mu}_1, \hat{\mu}_2, ..., \hat{\mu}_C, \hat{\eta}_1, \hat{\eta}_2, ..., \hat{\eta}_C]$ are the estimated parameters and $C$ is the corresponding number of components. The corresponding BIC score can be computed as

$$BIC_C = -2\, \mathbb{L}_C \left( \hat{\boldsymbol{\theta}} \right) + C \ln \left( \mathcal{M} \right). \quad (25)$$

It can be seen that the BIC penalizes the model complexity by adding the regularization coefficient, $C \ln(\mathcal{M})$. It is worth noting that although the EM algorithm maximizes the log-likelihood distribution, the BIC is an asymptotic approximation to the transformation of the Bayesian *a posteriori* probability, $\Pr(\hat{\boldsymbol{\theta}}|\boldsymbol{x}, C)$. As such, in a large-sample setting, the number of components determined by the BIC is asymptotically optimal from the perspective of the Bayesian posterior probability. Here, we select the candidate model satisfying the minimum BIC score, satisfying asymptotically the maximum Bayesian posterior probability as

$$C_{opt} = \arg\min_{C \in \mathcal{N}} BIC_C. \tag{26}$$

Fig. 1 depicts the normalized BIC versus the number of components for some fading scenarios selected from Section II. The corresponding optimal number of components, $C_{opt}$, indicated in the legend, will be adopted in the simulations and numerical results hereafter and will be denoted by $C$.

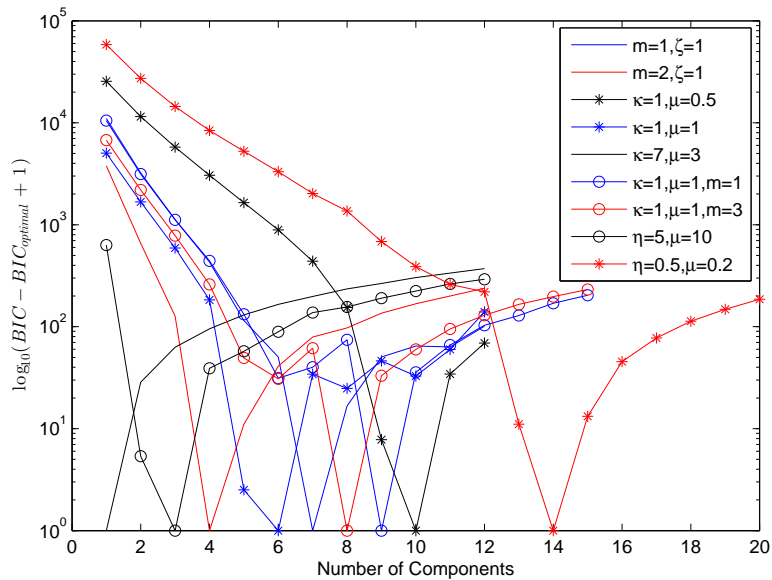

Figure 1.   Normalized BIC versus the number of components

Fig. 2 shows the optimal number of components as a function of the amount of fading. It is observed that as the fading becomes more severe, the mixture requires more components to accurately characterize the channel.

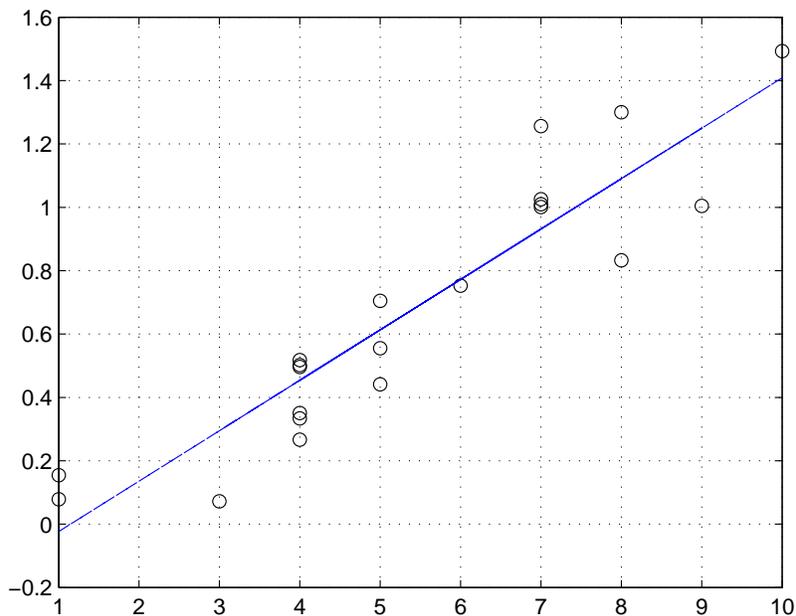

Figure 2. Optimal number of components versus the amount of fading

## IV. MoG Model Analysis and Comparisons

In this section, several scenarios of the aforementioned fading channels are approximated using the MoG distribution, as in (23). The number of components was selected automatically using the BIC method explained in Section III-C. We point out that higher accuracy can be achieved by increasing the number of components. In order to validate the accuracy of the approximations, we use two criteria of error, namely the MSE and the KL, defined as

$$\text{MSE} = \mathbb{E}\left[\hat{f}_\gamma(x) - f_\gamma(x)\right] \tag{27}$$

and

$$\text{KL}\left(f_\gamma(x), \hat{f}_\gamma(x)\right) = \int_0^\infty f_\gamma(x)\, \log \frac{f_\gamma(x)}{\hat{f}_\gamma(x)}\, dx, \tag{28}$$

respectively. Here, $f_\gamma(x)$ is the exact pdf, and $\hat{f}_\gamma(x)$ is the approximated pdf (MoG). The KL divergence, also known as relative entropy, is an information theoretic measure that quantifies

 

the information lost when $\hat{f}_\gamma(x)$ is used to approximate $f_\gamma(x)$ [36]. Note that the MSE and KL measures are used in several related works, see e.g., [8], [9], [12].

Fig. 3 provides the approximation results for several scenarios of the NL, $\kappa - \mu$, $\eta - \mu$, and $\kappa - \mu$ Shadowed fading models. The corresponding number of components are indicated in the legend. As shown from the MSE and KL measures, the approximation is very accurate when both increasing the shadowing and the multipath fading severity, whereas for large amount of fading, the number of components increases. For instance, for the $\eta - \mu$ distribution, when $\eta = 5$, and $\mu = 10$, the amount of fading and number of components were $0.0721$ and $3$, respectively, whereas when $\eta = 0.7$ and $\mu = 0.4$, the amount of fading and number of components increased to $1.2999$ and $8$, respectively. The parameters of the approximations are tabulated in Appendix A.

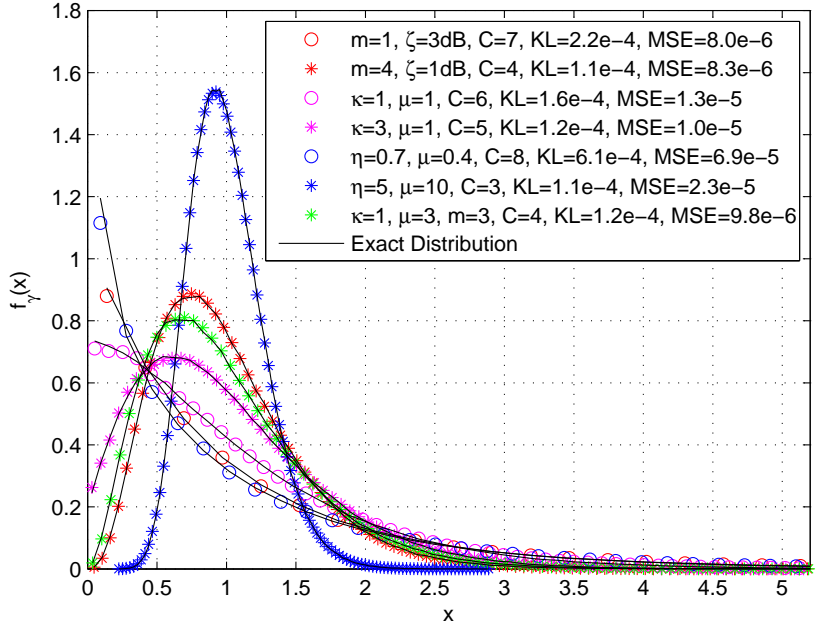

Figure 3.  MoG approximation for different channel models.

Further verification of the accuracy via numerical means is addressed in section VI. The purpose of this approach is not to increase the accuracy of the approximation, as it has already been achieved in all aforementioned fading alternatives, but rather to provide another unifying and simplifying distribution that has the potential of approximating all contemporary fading, composite and non-composite, models via the EM algorithm.

## V. Performance Analysis of Wireless Channels

The MoG distribution provides a simplifying and unifying analysis for wireless communication systems over various composite and multipath fading channel models. In this section, we first derive several performance metrics, which can be used for the evaluation of wireless communication systems in a generalized manner. In particular, we derive expressions for the raw moments of the MoG model, the amount of fading (AF), the outage probability, the channel capacity, and the MGF. We further derive expressions for the SER performance of $L$-branch MRC diversity system and the average probability of detection for cognitive radio networks.

### A. Moment Generating Function

By definition, the MGF $M_\gamma(s) = \mathbb{E}\left[e^{-s\gamma}\right]$ is given by

$$M_\gamma(s) = \sum_{i=1}^{C} \frac{\omega_i}{\sqrt{8\overline{\gamma}\pi}\eta_i} \int_0^\infty \frac{1}{\sqrt{\gamma}} \exp\left(-\frac{\left(\sqrt{\frac{2\gamma}{\overline{\gamma}}} - \mu_i\right)^2}{2\eta_i^2}\right) e^{-\gamma s}\, d\gamma. \tag{29}$$

Applying the change of variables $x = \sqrt{\frac{2\gamma}{\overline{\gamma}}}$, and after expanding the exponentials and considerable mathematical simplifications, we get

$$M_\gamma(s) = \sum_{i=1}^{C} \frac{\omega_i}{\sqrt{2\pi}\eta_i} \int_0^\infty \exp\left(-\frac{(2-\beta_i)\left(x^2 - \frac{2\mu_i}{\beta_i}x + \frac{\mu_i^2}{\beta_i}\right)}{2\eta_i^2}\right) dx, \tag{30}$$

where $\beta_i = 1 + 2\eta_i^2\overline{\gamma}s$. Then, after some mathematical manipulations, we obtain

$$M_\gamma(s) = \sum_{i=1}^{C} \frac{\omega_i \exp\left(\frac{\mu_i^2\left(\frac{1}{\beta_i}-1\right)}{2\eta_i^2}\right)}{\sqrt{\beta_i\pi}} \int_{\frac{-\mu_i}{\eta_i\sqrt{2\beta_i}}}^\infty \exp\left(-z^2\right) dz, \tag{31}$$

which leads to the following expression

$$M_\gamma(s) = \sum_{i=1}^{C} \frac{\omega_i}{\sqrt{\beta_i}} \exp\left(\frac{\mu_i^2 s}{\beta_i}\right) Q\left(-\frac{\mu_i}{\eta_i\sqrt{\beta_i}}\right),$$ (32)

where $Q(.)$ is the Gaussian Q-function defined as $Q(x) = \frac{1}{\sqrt{2\pi}} \int_x^\infty \exp\left(-\frac{u^2}{2}\right) du$.

## B. Raw Moments

The $n^{th}$ raw moment of the MoG distribution, by definition, is

$$\mathbb{E}[\gamma^n] = \sum_{i=1}^{C} \frac{\omega_i}{\sqrt{8\overline{\gamma}\pi}\eta_i} \int_0^\infty \frac{\gamma^n}{\sqrt{\gamma}} \exp\left(-\frac{\left(\sqrt{\frac{\gamma}{\overline{\gamma}}} - \mu_i\right)^2}{2\eta_i^2}\right) d\gamma.$$ (33)

By taking the change of variables $x = \sqrt{\frac{\gamma}{\overline{\gamma}}}$, and after some mathematical simplifications, we get

$$\mathbb{E}[\gamma^n] = \sum_{i=1}^{C} \omega_i \overline{\gamma}^n \int_0^\infty \frac{x^{2n}}{\sqrt{2\pi}\eta_i} \exp\left(-\frac{(x - \mu_i)^2}{2\eta_i^2}\right) dx,$$ (34)

Alternatively, we can write (33) as

$$\mathbb{E}[\gamma^n] = \sum_{i=1}^{C} \omega_i \overline{\gamma}^n \, \mathbb{E}[X_i^{2n}],$$ (35)

where $X_i \sim \mathcal{N}(\mu_i, \zeta_i)$ is the $i^{th}$ Gaussian component. Using the MGF approach, (35) can be expressed as

$$\mathbb{E}[\gamma^n] = \sum_{i=1}^{C} \omega_i \overline{\gamma}^n \frac{d^{(2n)} M_{X_i}(s)}{ds^{(2n)}}|_{s=0},$$ (36)

where $M_{X_i}(s) = \mathbb{E}\{e^{-sX_i}\}$ is the MGF of $X_i$, which is given by

$$M_{X_i}(s) = \exp\left(\mu_i s + \frac{\eta_i^2 s^2}{2}\right)$$ (37)

Equation (36) is mathematically convenient for solving the first few moments.

An alternative approach that yields a closed form expression can be attained by following the same method in [37], where the $v$th raw moments of $X_i$ are derived as

$$\mathbb{E}[X_i^v] = \eta_i^v 2^{\frac{v}{2}} \frac{\Gamma\left(\frac{v}{2} + \frac{1}{2}\right)}{\sqrt{\pi}} \, {}_1F_1\left[-\frac{v}{2}, \frac{1}{2}, -\frac{\mu_i^2}{2\eta_i^2}\right],$$ (38)

where $v$ is an even integer (note that there is no loss in generality).

By substituting (38) into (35), the $n^{th}$ raw moment of the MoG distribution is derived as

$$\mathbb{E}\left[\gamma^n\right] = \sum_{i=1}^{C} \omega_i \overline{\gamma}^n \eta_i^{2n} 2^n \frac{\Gamma\left(n + \frac{1}{2}\right)}{\sqrt{\pi}} {}_1F_1\left[-n, \frac{1}{2}, -\frac{\mu_i^2}{2\eta_i^2}\right],\tag{39}$$

## C. Amount of Fading

The AF measure was firstly introduced by Charash [38], as a measure of the severity of the fading channel. The AF requires the knowledge of only the first two moments in the corresponding fading channel, where it is defined by

$$\mathrm{AF} = \frac{\mathbb{E}\left[\gamma^2\right] - \mathbb{E}\left[\gamma\right]^2}{\mathbb{E}\left[\gamma\right]^2}.\tag{40}$$

By solving (39) for the first two moments, we obtain

$$\mathrm{AF} = \frac{\sum_{i=1}^{C} \omega_i \left(\mu_i^4 + 6\mu_i^2\eta_i^2 + 3\eta_i^4\right)}{\left[\sum_{i=1}^{C} \omega_i \left(\mu_i^2 + \eta_i^2\right)\right]^2} - 1.\tag{41}$$

## D. Outage Probability

The outage probability is a standard performance criterion used over fading channels. It is defined as $F(\gamma_{th}) = \int_0^{\gamma_{th}} f_\gamma(x)\,dx$. By performing the following change of variables applied to (23)

$$y = \frac{\sqrt{\frac{x}{\overline{\gamma}}} - \mu_i}{\eta_i},\tag{42}$$

and after some mathematical manipulations, the CDF of (23) can be written as

$$F\left(\gamma_{th}\right) = \sum_{i=1}^{C} \frac{\omega_i}{\sqrt{\pi}} \int_{\frac{-\mu_i}{\eta_i}}^{\frac{\sqrt{\frac{\gamma_{th}}{\overline{\gamma}}} - \mu_i}{\eta_i}} \exp\left(-\frac{y^2}{2}\right)\,dy.\tag{43}$$

Further simplifications yield

$$F\left(\gamma_{th}\right) = \sum_{i=1}^{C} \omega_i \left[Q\left(-\frac{\mu_i}{\eta_i}\right) - Q\left(\frac{\sqrt{\frac{\gamma_{th}}{\overline{\gamma}}} - \mu_i}{\eta_i}\right)\right],\tag{44}$$

## E. Average Ergodic Channel Capacity

When only the receiver has knowledge about the channel state information (CSI), the ergodic capacity $C_{erg}$ is expressed as

$$C_{erg} = \frac{B}{\ln 2} \int_0^\infty \ln\left(1 + \gamma\right) f_\gamma\left(\gamma\right) d\gamma. \tag{45}$$

where $B$ is the channel bandwidth measured in Hertz. Unfortunately, the exact solution of (45) is intractable. Instead, a computationally simple and very accurate form can be obtained by following [39], where $\ln(1 + \gamma)$ is expanded about the mean value of the instantaneous SNR, $\mathbb{E}[\gamma]$, using Taylor's series, yielding

$$\begin{aligned} \ln\left(1 + \gamma\right) &= \ln\left(1 + \mathbb{E}\left[\gamma\right]\right) + \sum_{w=1}^\infty \frac{(-1)^{w-1}}{w} \frac{\left(\gamma - \mathbb{E}\left[\gamma\right]\right)^w}{\left(1 + \mathbb{E}\left[\gamma\right]\right)^w} \\ &\approx \ln\left(1 + \mathbb{E}\left[\gamma\right]\right) + \frac{\gamma - \mathbb{E}\left[\gamma\right]}{1 + \mathbb{E}\left[\gamma\right]} + \frac{\left(\gamma - \mathbb{E}\left[\gamma\right]\right)^2}{2\left(1 + \mathbb{E}\left[\gamma\right]\right)^2} + o\left[\left(x - \mathbb{E}\left[\gamma\right]\right)^2\right]. \end{aligned} \tag{46}$$

where $o\left(.\right)$ is one of the Landau symbols defined as $f = o\left(\phi\right)$ means that $\frac{f}{\phi} \to 0$.

Substituting the logarithm function approximation from (46) into (45), a second order approximation for the ergodic capacity can be evaluated. It is noted that the solution to (45) is obtained by taking the expectation of $\ln\left(1 + \gamma\right)$. Hence, taking the expectation of (46), we get

$$C_{erg} \approx \frac{B}{\ln 2} \left[\ln\left(1 + \mathbb{E}\left[\gamma\right]\right) - \frac{\mathbb{E}\left[\gamma^2\right] - \mathbb{E}^2\left[\gamma\right]}{2\left(1 + \mathbb{E}\left[\gamma\right]\right)^2}\right], \tag{47}$$

where $\mathbb{E}\left[\gamma^n\right]$ is evaluated using (39).

## F. Symbol Error Analysis

In order to further demonstrate the significance of the MoG distribution, we study the performance of independent but not identically distributed ($i.n.i.d.$) $L$-branch MRC diversity receiver over various composite and non-composite fading scenarios. The MRC scheme is the optimal combining scheme at the expense of increased complexity, where the receiver requires knowledge of all channel fading parameters [2]. Here, the receiver sums up all received instantaneous SNR replicas $\gamma_k$ as follows

$$\gamma_{MRC} = \sum_{k=1}^{L} \gamma_k. \tag{48}$$

The corresponding MGF is thus

$$M_{\gamma_{MRC}}(s) = \mathbb{E}\{e^{-s\sum_{k=1}^{L}\gamma_k}\} = \prod_{k=1}^{L} M_{\gamma_k}(s), \tag{49}$$

where $M_{\gamma_k}(s)$ is derived in (32). The SER, $P_s(E)$, for coherent binary signals, can be computed as follows [2]

$$P_s(E) = \mathbb{E}_{\gamma_{MRC}}\left[Q\left(\sqrt{2g\gamma_{MRC}}\right)\right], \tag{50}$$

where $g$ is some constant resembling several coherent binary signals, such as coherent binary phase shift keying (BPSK) and coherent orthogonal binary frequency shift keying (BFSK) corresponding to $g = 1$ and $g = \frac{1}{2}$, respectively. By substituting the Q-function by its definition in [2, eq. 4.2], the SER is written as

$$P_s(E) = \frac{1}{\pi} \int_0^{\frac{\pi}{2}} \int_0^{\infty} \exp\left(-\frac{g\,\gamma_{MRC}}{\sin^2(\theta)}\right) f_{\gamma_{MRC}}(\gamma_{MRC})\, d\gamma_{MRC}\, d\theta. \tag{51}$$

The inner infinite integral in (51) is the equivalent MGF derived in (49), yielding

$$P_s(E) = \frac{1}{\pi} \int_0^{\frac{\pi}{2}} \prod_{k=1}^{L} M_{\gamma_k}\left(\frac{g}{\sin^2(\theta)}\right) d\theta, \tag{52}$$

where $M_{\gamma_k}(.)$ was derived in (32). Following a similar approach, and by utilizing [2, eq. 8.23] and [2, eq. 8.12], the SER expressions for $M$-PSK and square $M$-QAM signaling schemes are given, respectively, by

$$P_s(E) = \frac{1}{\pi} \int_0^{\frac{(M-1)\pi}{M}} \prod_{k=1}^{L} M_{\gamma_k}\left(\frac{\sin^2\left(\frac{\pi}{M}\right)}{\sin^2(\theta)}\right) d\theta, \tag{53}$$

$$P_s(E) = \frac{4}{\pi}\left(\frac{\sqrt{M}-1}{\sqrt{M}}\right)\left[\int_0^{\frac{\pi}{2}} \prod_{k=1}^{L} M_{\gamma_k}\left(\frac{g_{QAM}}{\sin^2(\theta)}\right) d\theta - \left(\frac{\sqrt{M}-1}{\sqrt{M}}\right)\int_0^{\frac{\pi}{4}} \prod_{k=1}^{L} M_{\gamma_k}\left(\frac{g_{QAM}}{\sin^2(\theta)}\right) d\theta\right], \tag{54}$$

where $g_{QAM} = \frac{3}{2}(M-1)$.

## G. Probability of detection

Cognitive radio (CR) is a promising technology that can enhance the performance of wireless communications [40]. The basic concept behind opportunistic CR, is that a secondary user is allowed to use the spectrum, which is assigned to a licensed primary user (PU), when the channel is idle [41]. The CR users perform spectrum sensing in order to identify idle spectrum. Energy detection is the most common sensing technique in CR networks, due to its low implementation complexity and no requirements for knowledge of the signal [42]. Several studies have been devoted to the analysis of the performance of energy detection-based spectrum sensing for different communication and fading scenarios [43]. The probability of detection, $P_d$ is an important performance evaluation metric representing the probability that an active PU is detected by the CR node. Over fading channels, the average detection probability is evaluated by averaging the AWGN channel $P_d$ over the SNR distribution as follows:

$$\overline{P_d} = \int_0^\infty Q_u \left( \sqrt{2\gamma}, \sqrt{\lambda} \right) f_\gamma(\gamma) \, d\gamma, \tag{55}$$

where $Q_u \left( \sqrt{2\gamma}, \sqrt{\lambda} \right)$ is the probability of detection over AWGN channel with $Q_u(.,.)$ being the generalized Marcum-Q function [44]. Moreover, $\lambda$ is a predefined energy detection threshold, $u$ is the time bandwidth product which corresponds to the number of samples of either the in-phase ($I$) or the quadrature ($Q$) component, and $\gamma \triangleq \frac{\alpha^2 E_s}{N_0}$ is the received SNR, where $E_s$ is the signal energy, $N_0$ the one-sided noise power spectral density, and the channel gain $\mathbb{E}\left[|\alpha^2|\right] = 1$.

The Marcum Q-function can be expressed as [45]

$$Q_u \left( \sqrt{2\gamma}, \sqrt{\lambda} \right) = e^{-\gamma} \sum_{n=0}^\infty \frac{\gamma^n}{n!} \frac{\Gamma\left(u+n, \frac{\lambda}{2}\right)}{\Gamma(u+n)}. \tag{56}$$

By substituting (23) and (56) in (55), we obtain

$$\overline{P_d} \simeq \sum_{i=1}^C \frac{w_i e^{-\frac{\mu_i^2}{2\eta_i^2}\left(1 - \frac{1}{4\overline{\gamma}\eta_i^2 + 2}\right)}}{\sqrt{2\pi}\overline{\gamma}\eta_i} \sum_{n=0}^\infty \frac{\Gamma\left(u+n, \frac{\lambda}{2}\right)\Gamma(2n+1)\left(\overline{\gamma}\eta_i^2\right)^{\frac{2n+1}{2}}}{n!\Gamma(u+n)\left(2\overline{\gamma}\eta_i^2 + 1\right)^{\frac{2n+1}{2}}} D_{-(2n+1)}\left(\frac{-\mu_i}{\eta_i\sqrt{2\overline{\gamma}\eta_i^2 + 1}}\right) \tag{57}$$

 

where $D_n(i)$ is the parabolic cylinder function [16].

The truncation of the infinite series in (57) generates the following error

$$\epsilon_t = \sum_{i=1}^{C} \frac{w_i e^{-\frac{\mu_i^2}{2\eta_i^2}\left(1-\frac{1}{4\overline{\gamma}\eta_i^2+2}\right)}}{\sqrt{2\pi\overline{\gamma}}\eta_i} \sum_{n=p+1}^{\infty} \frac{\Gamma\left(u+n,\frac{\lambda}{2}\right)\Gamma(2n+1)(\overline{\gamma}\eta_i^2)^{\frac{2n+1}{2}}}{n!\Gamma(u+n)(2\overline{\gamma}\eta_i^2+1)^{\frac{2n+1}{2}}} D_{-(2n+1)}\left(-\frac{\mu_i}{\eta_i\sqrt{2\overline{\gamma}\eta_i^2+1}}\right) \quad (58)$$

$$= \sum_{i=1}^{C} \frac{w_i e^{-\frac{\mu_i^2}{2\eta_i^2}\left(1-\frac{1}{4\overline{\gamma}\eta_i^2+2}\right)}}{\sqrt{2\pi\overline{\gamma}}\eta_i} \left[ \sum_{n=0}^{\infty} \frac{\Gamma\left(u+n,\frac{\lambda}{2}\right)\Gamma(2n+1)(\overline{\gamma}\eta_i^2)^{\frac{2n+1}{2}}}{n!\Gamma(u+n)(2\overline{\gamma}\eta_i^2+1)^{\frac{2n+1}{2}}} D_{-(2n+1)}\left(-\frac{\mu_i}{\eta_i\sqrt{2\overline{\gamma}\eta_i^2+1}}\right) \right.$$

$$\left. - \sum_{n=0}^{p} \frac{\Gamma\left(u+n,\frac{\lambda}{2}\right)\Gamma(2n+1)(\overline{\gamma}\eta_i^2)^{\frac{2n+1}{2}}}{n!\Gamma(u+n)(2\overline{\gamma}\eta_i^2+1)^{\frac{2n+1}{2}}} D_{-(2n+1)}\left(-\frac{\mu_i}{\eta_i\sqrt{2\overline{\gamma}\eta_i^2+1}}\right) \right] \quad (59)$$

Since the $\Gamma(a,x)$ function is monotonically decreasing with respect to $x$, thus, $\Gamma\left(u+n,\frac{\lambda}{2}\right) \leq \Gamma(u+n)$. Hence, the infinite sum in (59) can be upper bounded by:

$$\tau = \sum_{n=0}^{\infty} \frac{\Gamma(2n+1)}{n!} \left(\frac{\overline{\gamma}\eta_i^2}{2\overline{\gamma}\eta_i^2+1}\right)^{\frac{2n+1}{2}} D_{-(2n+1)}\left(-\frac{\mu_i}{\eta_i\sqrt{2\overline{\gamma}\eta_i^2+1}}\right) \quad (60)$$

By expressing the parabolic cylinder function in terms of the confluent hypergeometric function according to [16, eq. (9.240)], one obtains

$$\tau = \sqrt{\pi} e^{-\frac{\mu_i^2}{4\eta_i^2(2\overline{\gamma}\eta_i^2+1)}} \sum_{n=0}^{\infty} \frac{\Gamma(2n+1)}{n!2^{\frac{2n+1}{2}}} \left(\frac{\overline{\gamma}\eta_i^2}{2\overline{\gamma}\eta_i^2+1}\right)^{\frac{2n+1}{2}}$$

$$\times \left[ \frac{1}{n!} {}_1F_1\left(n+\frac{1}{2},\frac{1}{2},\frac{\mu_i^2}{2\eta_i^2(2\overline{\gamma}\eta_i^2+1)}\right) + \frac{\sqrt{2}\mu_i}{\Gamma\left(n+\frac{1}{2}\right)\eta_i\sqrt{2\overline{\gamma}\eta_i^2+1}} {}_1F_1\left(n+1,\frac{3}{2},\frac{\mu_i^2}{2\eta_i^2(2\overline{\gamma}\eta_i^2+1)}\right) \right] \quad (61)$$

By expanding the involved hypergeometric functions and after some algebraic manipulations, the following equality is valid

$$\tau = \sqrt{\pi} e^{-\frac{\mu_i^2}{4\eta_i^2(2\overline{\gamma}\eta_i^2+1)}} \sum_{n=0}^{\infty} \sum_{i=0}^{\infty} \frac{\left(n+\frac{1}{2}\right)_i \left(\frac{\mu_i^2}{2\eta_i^2(2\overline{\gamma}\eta_i^2+1)}\right)^i}{\left(\frac{1}{2}\right)_i} \frac{(n+2)\Gamma(n+2)}{n!} \left(\frac{\overline{\gamma}\eta_i^2}{4\overline{\gamma}\eta_i^2+2}\right)^{\frac{2n+1}{2}}$$

$$+ \frac{\mu_i\sqrt{2\pi}}{\eta_i\sqrt{2\overline{\gamma}\eta_i^2+1}} e^{-\frac{\mu_i^2}{4\eta_i^2(2\overline{\gamma}\eta_i^2+1)}} \sum_{n=0}^{\infty} \sum_{i=0}^{\infty} \frac{(n+1)_i \left(\frac{\mu_i^2}{2\eta_i^2(2\overline{\gamma}\eta_i^2+1)}\right)^i}{\left(\frac{3}{2}\right)_i} \frac{(n+2)\Gamma(n+2)}{n!} \left(\frac{\overline{\gamma}\eta_i^2}{4\overline{\gamma}\eta_i^2+2}\right)^{\frac{2n+1}{2}} \quad (62)$$

By recalling that $x = x!/(x-1)!$ and with the aid of the Pochhammer symbol identities, it follows that

$$\tau = \sqrt{\frac{\overline{\gamma}\pi\eta_i^2}{4\overline{\gamma}\eta_i^2+2}}\, e^{-\frac{\mu_i^2}{4\eta_i^2(2\overline{\gamma}\eta_i^2+1)}}\left[\sum_{n=0}^{\infty}\sum_{i=0}^{\infty}\frac{\left(\frac{1}{2}\right)_{n+i}(3)_n}{\left(\frac{1}{2}\right)_i(1)_n}\frac{\left(\frac{\overline{\gamma}\eta_i^2}{4\overline{\gamma}\eta_i^2+2}\right)^n}{n!}\frac{\left(\frac{\mu_i^2}{2\eta_i^2(2\overline{\gamma}\eta_i^2+1)}\right)^i}{i!}\right.$$

$$\left. +\frac{\mu_i\sqrt{2}}{\eta_i\sqrt{2\overline{\gamma}\eta_i^2+1}}\sum_{n=0}^{\infty}\sum_{i=0}^{\infty}\frac{(1)_{n+i}(3)_n}{\left(\frac{3}{2}\right)_i(1)_n}\frac{\left(\frac{\overline{\gamma}\eta_i^2}{4\overline{\gamma}\eta_i^2+2}\right)^n}{n!}\frac{\left(\frac{\mu_i^2}{2\eta_i^2(2\overline{\gamma}\eta_i^2+1)}\right)^i}{i!}\right] \tag{63}$$

Importantly, the above infinite series can be expressed in closed-form in terms of the Humbert $\Psi_1$ function [46] yielding

$$\epsilon_t < \sum_{i=1}^{C}\frac{w_i e^{-\frac{\mu_i^2}{2\eta_i^2}}}{\sqrt{2\pi\overline{\gamma}}\eta_i}\left[\sqrt{\frac{\overline{\gamma}\pi\eta_i^2}{4\overline{\gamma}\eta_i^2+2}}\Psi_1\left(\frac{1}{2},3,\frac{1}{2},1;\frac{\overline{\gamma}\pi\eta_i^2}{4\overline{\gamma}\eta_i^2+2},\frac{\mu_i^2}{2\eta_i^2(2\overline{\gamma}\eta_i^2+1)}\right)\right.$$

$$+\mu_i\sqrt{\overline{\gamma}\pi}\Psi_1\left(1,3,\frac{3}{2},1;\frac{\overline{\gamma}\pi\eta_i^2}{4\overline{\gamma}\eta_i^2+2},\frac{\mu_i^2}{2\eta_i^2(2\overline{\gamma}\eta_i^2+1)}\right)$$

$$\left. -e^{\frac{\mu_i^2}{4\eta_i^2(2\overline{\gamma}\eta_i^2+1)}}\sum_{n=0}^{p}\frac{\Gamma\left(u+n,\frac{\lambda}{2}\right)\Gamma(2n+1)(\overline{\gamma}\eta_i^2)^{\frac{2n+1}{2}}}{n!\Gamma(u+n)(2\overline{\gamma}\eta_i^2+1)^{\frac{2n+1}{2}}}D_{-(2n+1)}\left(\frac{-\mu_i}{\eta_i\sqrt{2\overline{\gamma}\eta_i^2+1}}\right)\right] \tag{64}$$

## VI. SIMULATION RESULTS

In this section, we present some analytical and simulation results for the outage probability, the average ergodic capacity, the SER of MRC scheme and the average detection probability for cognitive radio.

Fig. 4 and 5 depict the outage probability, as in (44), versus the threshold SNR $\gamma_{th}$ and the average SNR $\overline{\gamma}$, respectively. Two NL scenarios are considered in Fig. 4, where the multipath severity is reduced from $m = 1$ to $m = 3$ , and two $\kappa - \mu$ scenarios are considered in Fig.

5, where the control parameter $\kappa$ is increased from $\kappa = 1$ to $\kappa = 3$. Here, one can notice how accurate the approximation is. Also, it is very noticeable how the multipath fading severity affects the outage probability performance.

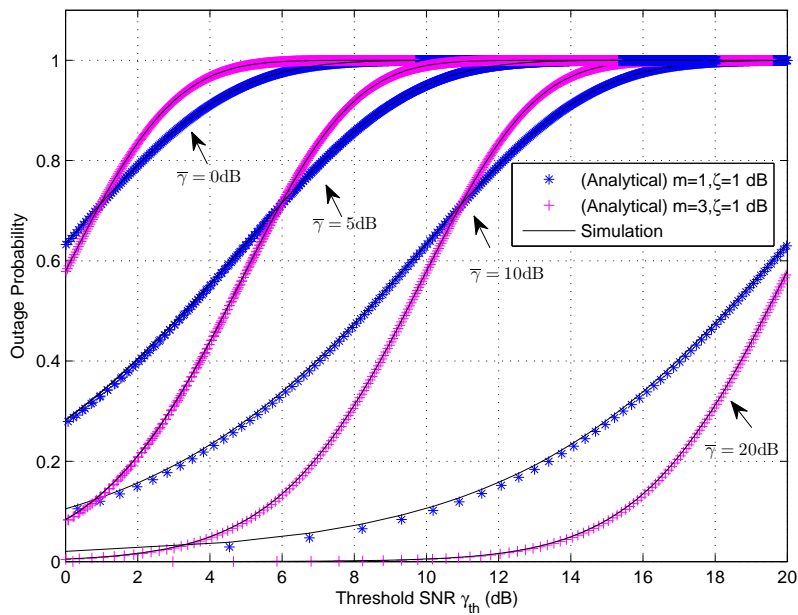

Figure 4. Analytical and simulated outage probability versus $\gamma_{th}$ for two scenarios.

Fig. 6 depicts the capacity, as in (47), for some selected scenarios from the $\eta - \mu$, $\kappa - \mu$ and $\kappa - \mu$ Shadowed fading models. As shown, the severe NLOS configuration of the $\eta - \mu$ scenario exhibits the worse capacity. In addition, one can see that introducing the Nakagami-$m$ shadowing to the $\kappa - \mu$ distribution has worsened the capacity. The term simulation refers to cross-validating the results via the recursive adaptive simpson quadrature method performed by the aid of a mathematical package.

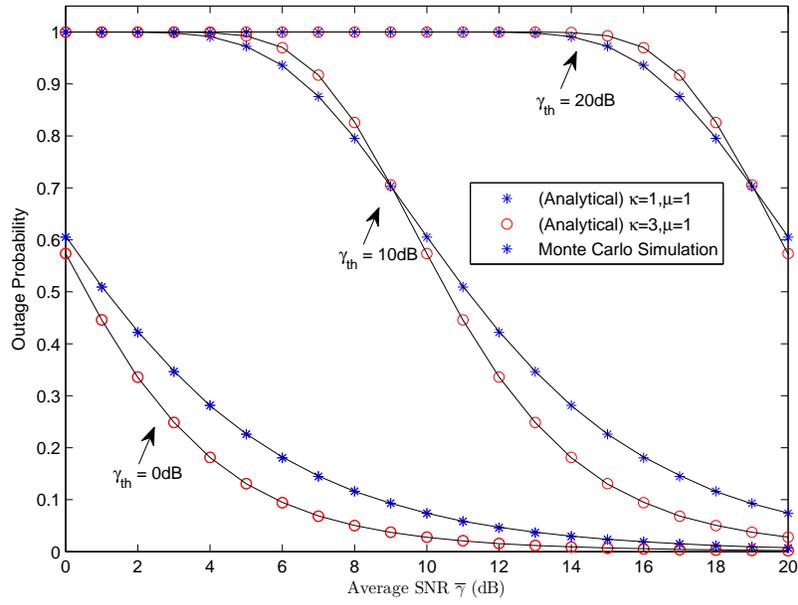

Figure 5.   Analytical and simulated outage probability versus $\overline{\gamma}$ for two scenarios.

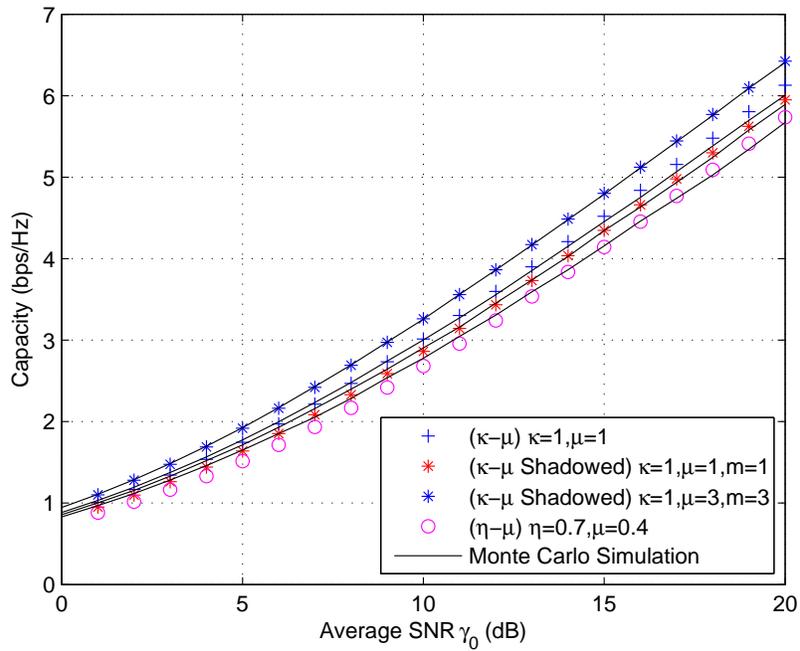

Figure 6.   Analytical and simulated ergodic capacity, $B=\frac{1}{2}$.

Fig. 7 illustrates the analytical SER of the BPSK signaling scheme for various NL scenarios, including mild and severe fading cases in both shadowing and multipath. The solid squared line represents the corresponding Monte Carlo simulation. It is quite noticeable how the multipath severity plays greater role in determining the SER, where as observed, incrementing $m$ by only 1 yields a SER performance improvement of about an order of magnitude at observed mid-range average SNR values. On the other hand, increasing $\zeta$ from 1 to 3 dB, while fixing $m$, yields a very similar SER performance.

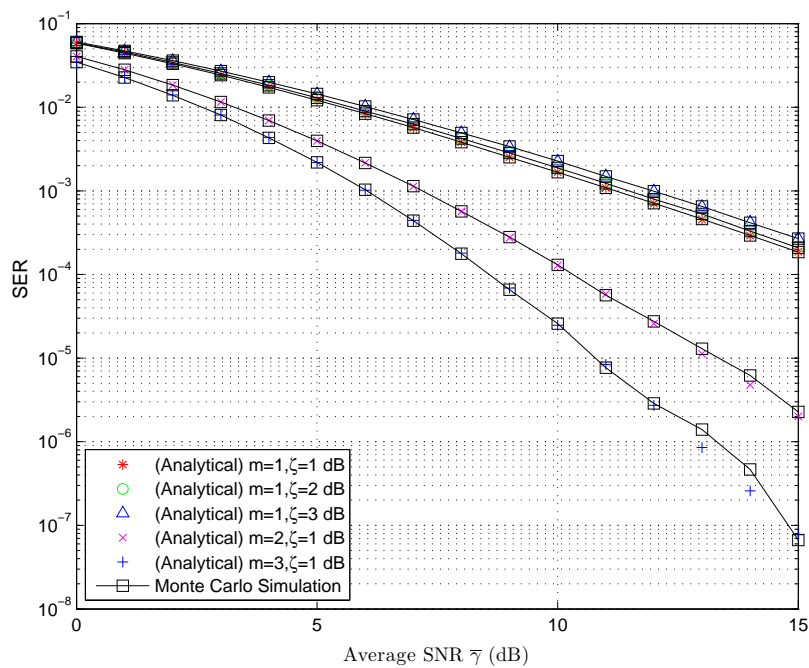

Figure 7. Analytical and simulation SER of 2-branch MRC diversity receiver for BPSK signaling scheme for RL and NL fading channels.

Fig. 8 features the analytical SER of the 16-QAM signaling scheme for various $\kappa-\mu$ Shadowed fading scenarios, where the control parameter $\mu$ and the shadowing severity parameter $m$ are varied. Here, $L$ corresponds to the number of antennas, and it is noticeable that our model is still very accurate for high antenna diversity order.

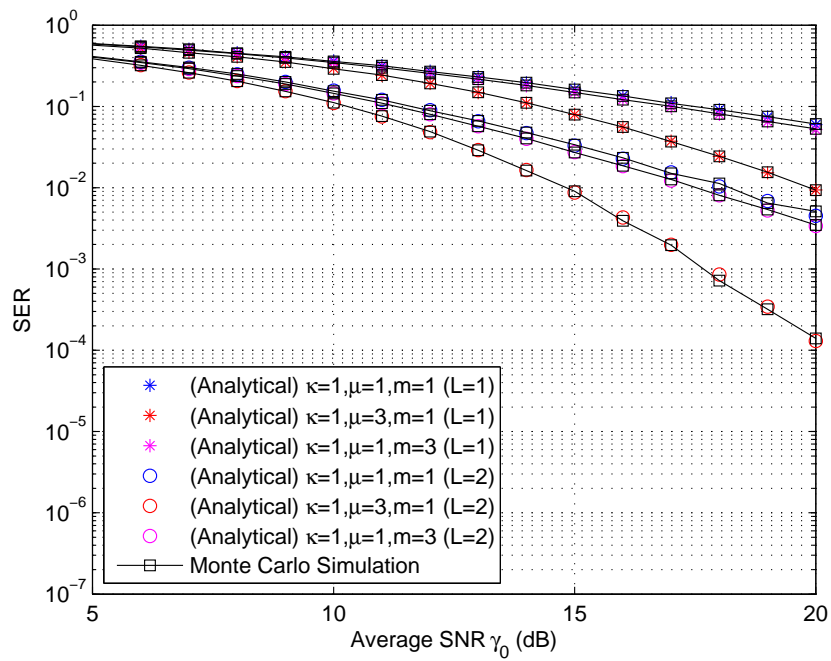

Figure 8. Analytical and simulation SER of $L$-branch MRC diversity receiver for 16-QAM signaling scheme for for various $\kappa - \mu$ Shadowed fading scenarios.

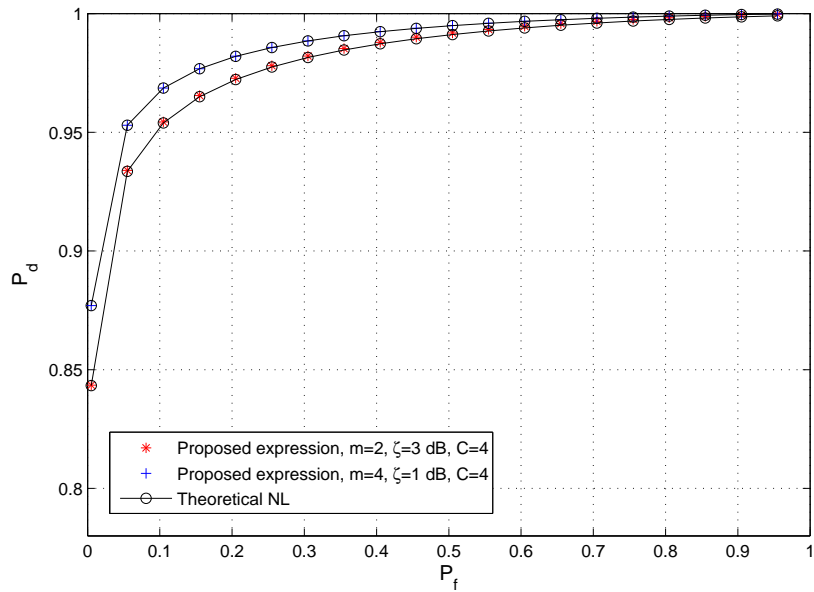

Figure 9. ROC curves for Nakagami-Lognormal channel with $u = 3$ and $\overline{\gamma} = 5$ dB

Fig. 9 depicts the receiver operating characteristic (ROC) curve, where $\overline{P_d}$ evaluated using (57) is compared to the theoretical expression from (55) substituting $f_\gamma(\gamma)$ by the NL fading channel pdf. These curves, depicting $\overline{P_d}$ versus the probability of false alarm ($P_f$), are essential for the performance evaluation of the CR. We consider a single user performing spectrum sensing over NL channel where $m$ is the Nakagami-$m$ fading parameter and $\zeta^2$, measured in dB, is the variance of the Gaussian random variable defined by $V = 10log_{10}(\sigma)$, where $\sigma$ corresponds to the Lognormal shadowing. The parameters for the MoG distribution are calculated via the EM algorithm and are provided in Appendix A. It can be observed that, for both cases, the approximation is very accurate over all the values of ($P_f$). Moreover, by evaluating the truncation error upper bound using (64), we conclude that although the truncation error depends on many parameters, considering the practical case where $P_f = 0.1$ and truncating the series at $n = 12$ ensures that the error does not exceed $0.08$ for the considered channels.

## VII. Conclusion

In this paper, the MoG distribution has been considered to characterize the amplitude and the SNR statistics for wireless propagation. The parameters of the mixtures are evaluated by the means of the EM algorithm and the MSE and KL have been evaluated to challenge the proposed model's accuracy. The proposed distribution enjoys both simplicity and accuracy and we have shown that the proposed pdf expression can accurately represent a wide range of both composite and non-composite channels. It should be highlighted that the adopted approach provides a generalized distribution for wireless communication systems where all channels can be modeled with the same analytical expression. Several analytical tools essential for the evaluation of performance analysis of digital communications were presented. This new model can be applied to various scenarios including, diversity systems, cooperative communications, and cognitive radio networks.

APPENDIX A

MoG PARAMETERS FOR SELECTED SCENARIOS

The following tables provide the approximation parameters for all scenarios presented in the paper.

Table I

MoG PARAMETERS FOR RL FADING CHANNEL WITH $\zeta = 3$ dB AND $C_{opt} = 7$

| $i$ | $w_i$ | $\mu_i$ | $\eta_i$ |
|---|---|---|---|
| 1 | 0.24621 | 0.76637 | 0.1888 |
| 2 | 0.28164 | 1.0954 | 0.27482 |
| 3 | 0.15143 | 1.5254 | 0.39455 |
| 4 | 0.077823 | 0.26911 | 0.088815 |
| 5 | 0.025355 | 2.0441 | 0.59471 |
| 6 | 0.19662 | 0.48573 | 0.14066 |
| 7 | 0.020909 | 0.11846 | 0.050183 |

Table II

MoG PARAMETERS FOR NL FADING CHANNEL WITH $m = 2$, $\zeta = 1$ dB AND $C_{opt} = 4$

| $i$ | $w_i$ | $\mu_i$ | $\eta_i$ |
|---|---|---|---|
| 1 | 0.11009 | 0.48576 | 0.14483 |
| 2 | 0.14047 | 1.3455 | 0.34267 |
| 3 | 0.37935 | 1.0845 | 0.25099 |
| 4 | 0.37009 | 0.77746 | 0.19545 |

 

Table III

MoG Parameters for NL Fading Channel with $m = 4$, $\zeta = 1\,\mathrm{dB}$ and $C_{opt} = 4$

| $i$ | $w_i$ | $\mu_i$ | $\eta_i$ |
|---|---|---|---|
| 1 | 0.31126 | 0.88351 | 0.14927 |
| 2 | 0.13366 | 1.2306 | 0.25803 |
| 3 | 0.39008 | 1.0756 | 0.19479 |
| 4 | 0.165 | 0.67541 | 0.1405 |

Table IV

MoG Parameters for $\kappa - \mu$ Fading Channel with $\kappa = 1$, $\mu = 0.5$ and $C_{opt} = 10$

| $i$ | $w_i$ | $\mu_i$ | $\eta_i$ | $i$ | $w_i$ | $\mu_i$ | $\eta_i$ |
|---|---|---|---|---|---|---|---|
| 1 | 0.005263 | 0.0052273 | 0.0032608 | 6 | 0.1401 | 0.39636 | 0.11622 |
| 2 | 0.050502 | 0.11458 | 0.038823 | 7 | 0.013249 | 0.020352 | 0.0089878 |
| 3 | 0.18712 | 1.4475 | 0.3582 | 8 | 0.22111 | 0.6682 | 0.18341 |
| 4 | 0.21379 | 1.0393 | 0.24872 | 9 | 0.026951 | 0.052855 | 0.019739 |
| 5 | 0.054249 | 1.8764 | 0.48797 | 10 | 0.087675 | 0.22199 | 0.069679 |

Table V

MoG Parameters for $\kappa - \mu$ Fading Channel with $\kappa = 3$, $\mu = 1$ and $C_{opt} = 5$

| $i$ | $w_i$ | $\mu_i$ | $\eta_i$ |
|---|---|---|---|
| 1 | 0.248 | 1.2295 | 0.29068 |
| 2 | 0.23197 | 0.59067 | 0.17727 |
| 3 | 0.017972 | 0.24425 | 0.099698 |
| 4 | 0.2658 | 1.0976 | 0.20736 |
| 5 | 0.23626 | 0.86379 | 0.16438 |

Table VI

MoG Parameters for $\eta - \mu$ Fading Channel with $\eta = 0.5$, $\mu = 0.2$ and $C_{opt} = 14$

| $i$ | $w_i$ | $\mu_i$ | $\eta_i$ | $i$ | $w_i$ | $\mu_i$ | $\eta_i$ |
|---|---|---|---|---|---|---|---|
| 1 | 0.092351 | 0.21051 | 0.061239 | 8 | 0.17858 | 0.58555 | 0.15984 |
| 2 | 0.030907 | 0.034416 | 0.012126 | 9 | 0.0052011 | 0.00092585 | 0.00068077 |
| 3 | 0.063814 | 0.12078 | 0.036005 | 10 | 0.0091546 | 0.0047069 | 0.0022762 |
| 4 | 0.092301 | 1.8163 | 0.41813 | 11 | 0.11756 | 1.352 | 0.28677 |
| 5 | 0.00030556 | 4.5372 | 0.47284 | 12 | 0.019634 | 0.014034 | 0.0055603 |
| 6 | 0.1304 | 0.35591 | 0.10026 | 13 | 0.029039 | 2.5233 | 0.61922 |
| 7 | 0.18817 | 0.92188 | 0.22362 | 14 | 0.042591 | 0.067048 | 0.021155 |

Table VII

MoG Parameters for $\eta - \mu$ Fading Channel with $\eta = 5$, $\mu = 10$ and $C_{opt} = 3$

| $i$ | $w_i$ | $\mu_i$ | $\eta_i$ |
|---|---|---|---|
| 1 | 0.32677 | 1.063 | 0.13602 |
| 2 | 0.32789 | 0.89548 | 0.099855 |
| 3 | 0.34533 | 1.0116 | 0.10407 |

Table VIII

MoG Parameters for $\kappa - \mu$ Shadowed Fading Channel with $\kappa = 1$, $\mu = 3$, $m = 3$ and $C_{opt} = 4$

| $i$ | $w_i$ | $\mu_i$ | $\eta_i$ |
|---|---|---|---|
| 1 | 0.40871 | 0.86626 | 0.18319 |
| 2 | 0.40942 | 1.111 | 0.22912 |
| 3 | 0.049081 | 1.4037 | 0.25915 |
| 4 | 0.13279 | 0.61749 | 0.15039 |

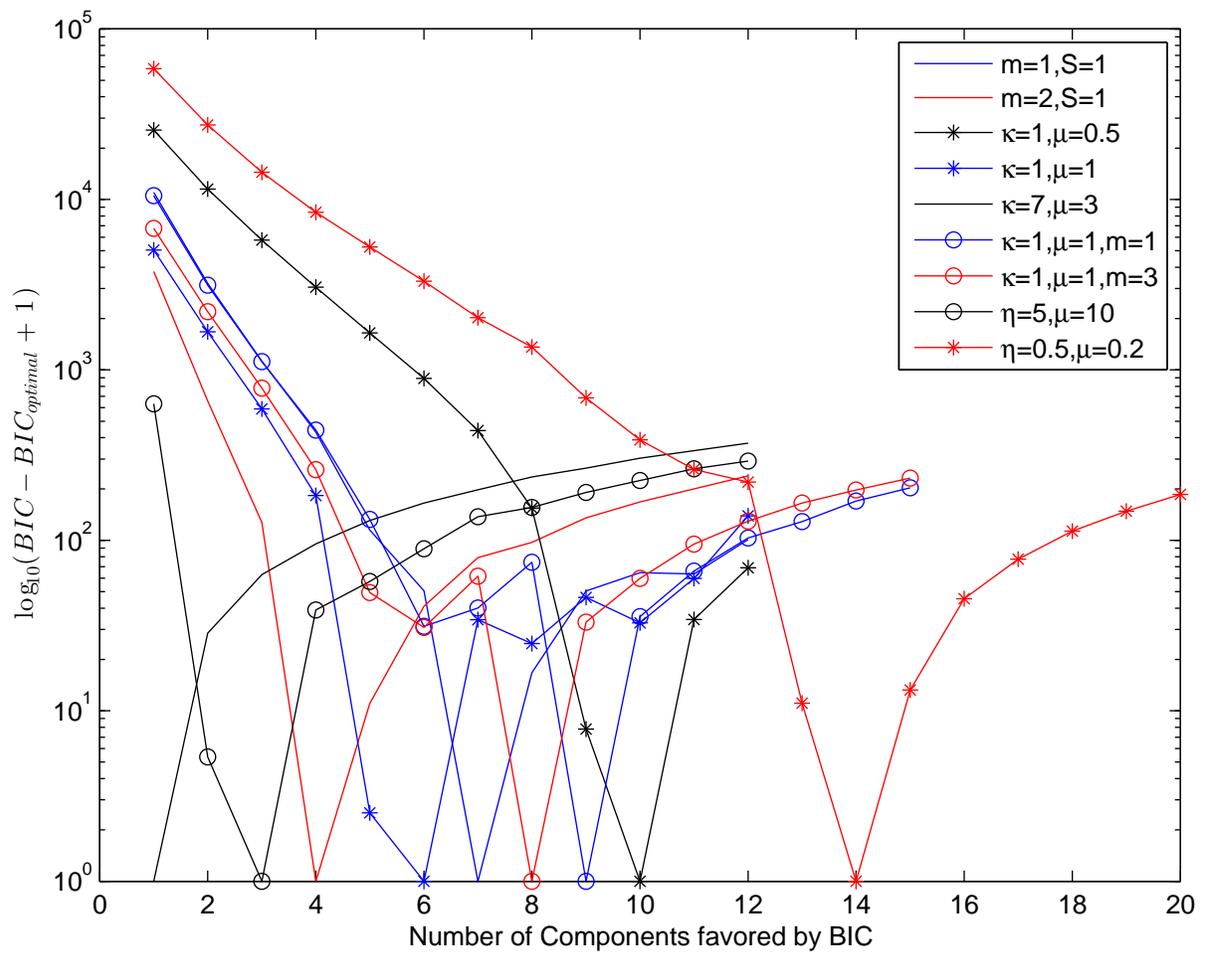

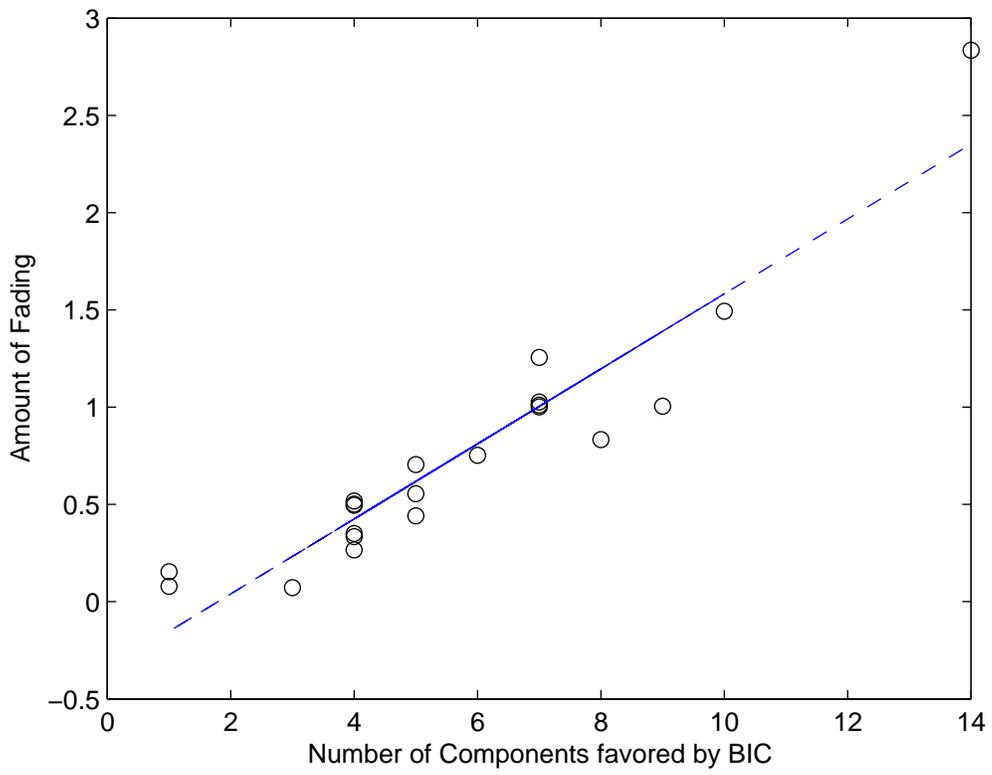

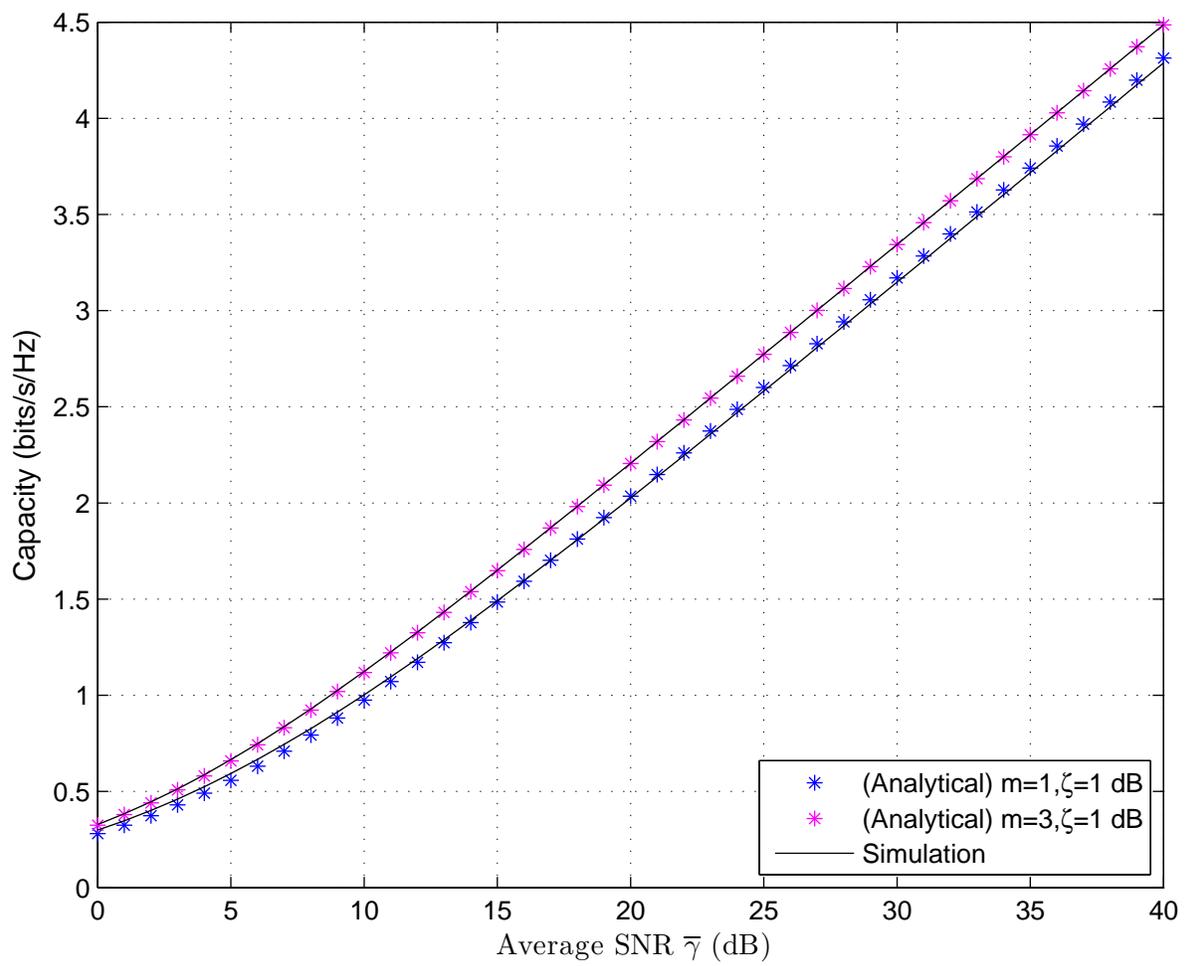

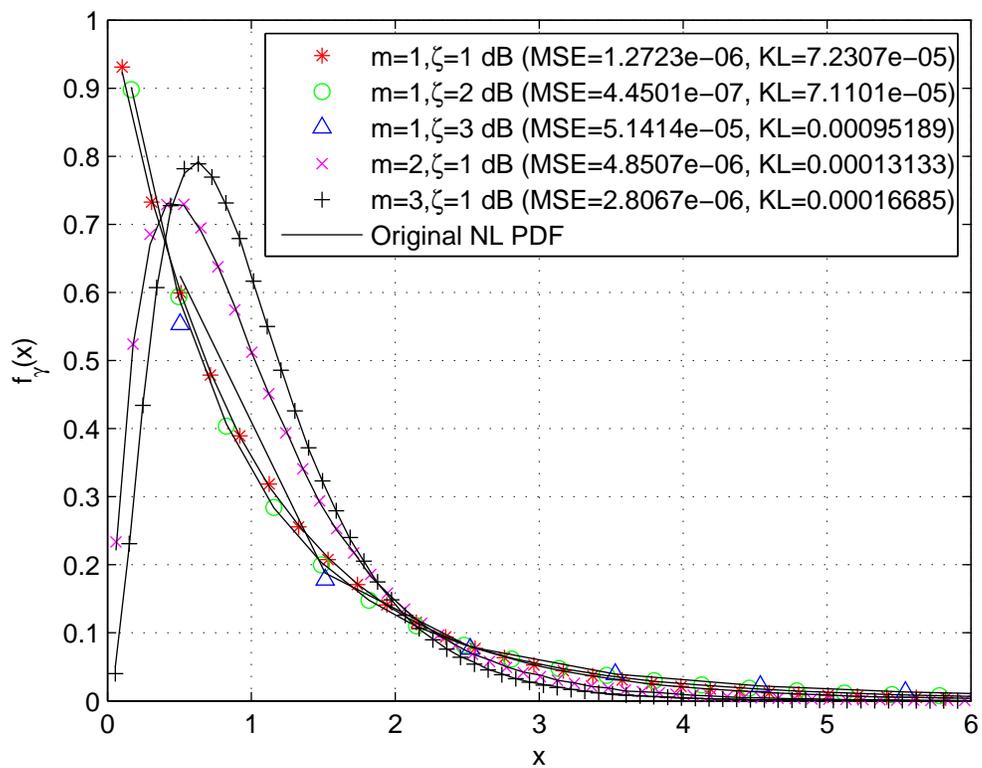

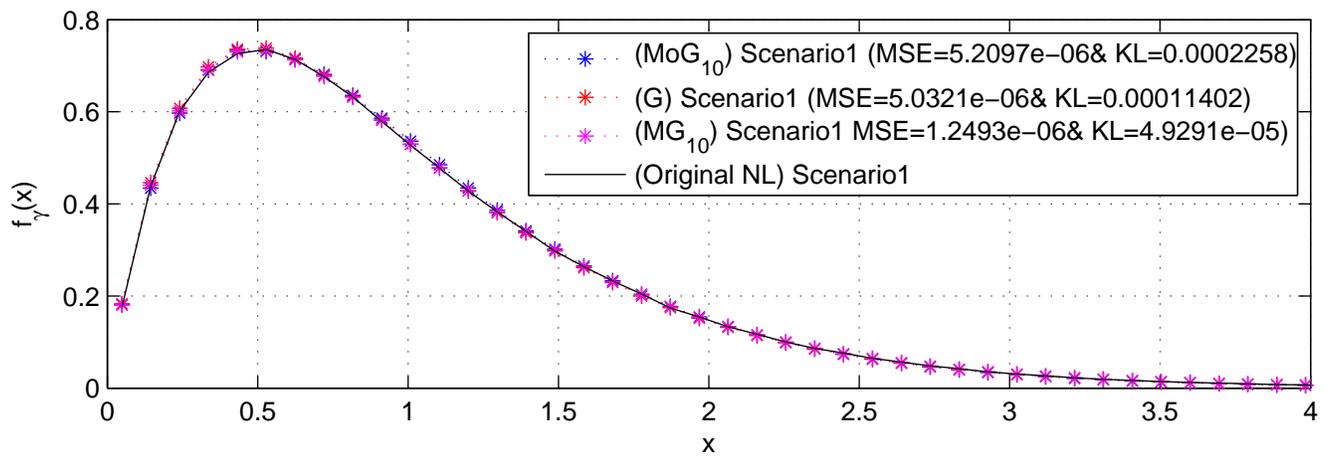

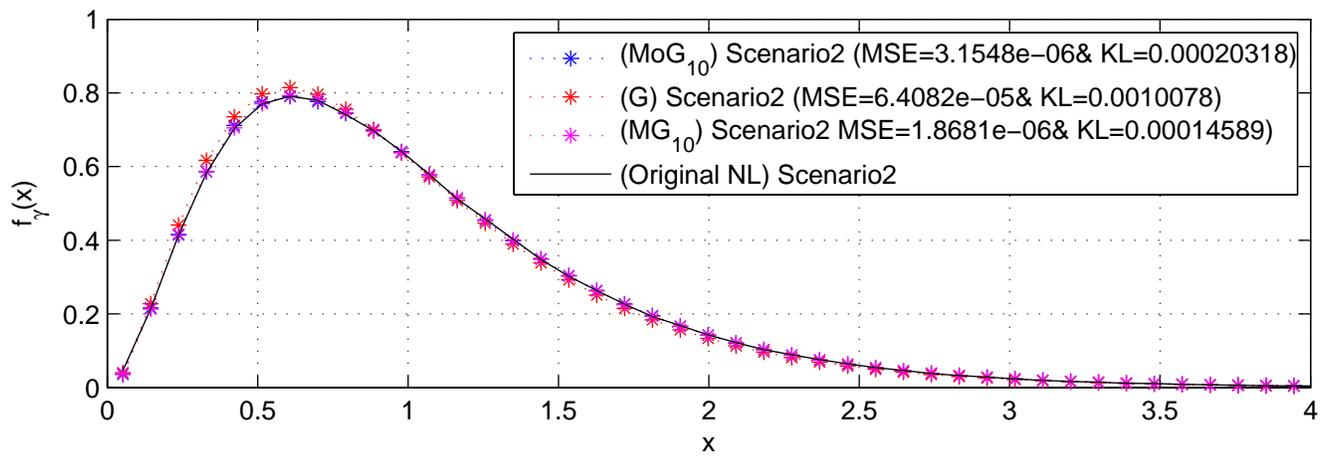

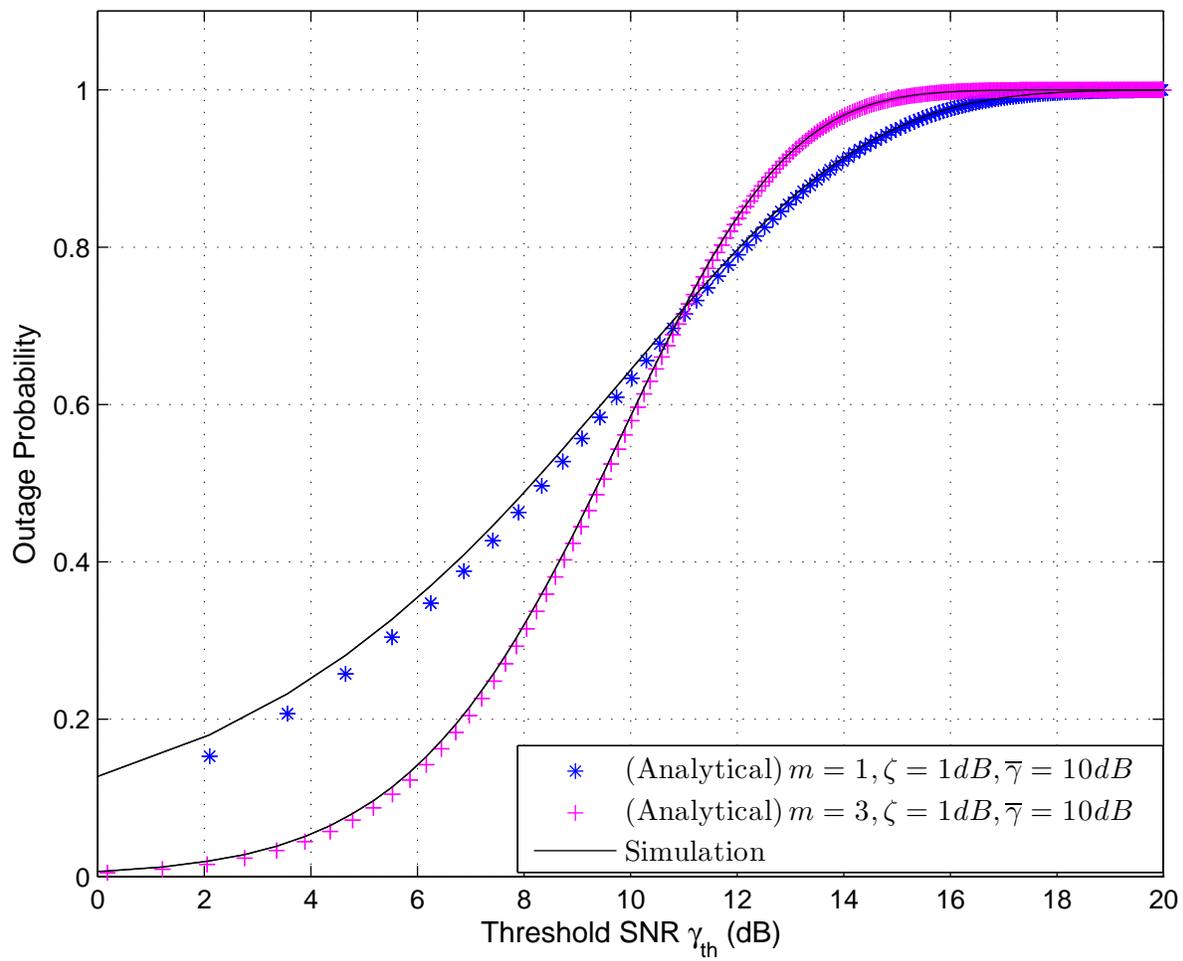

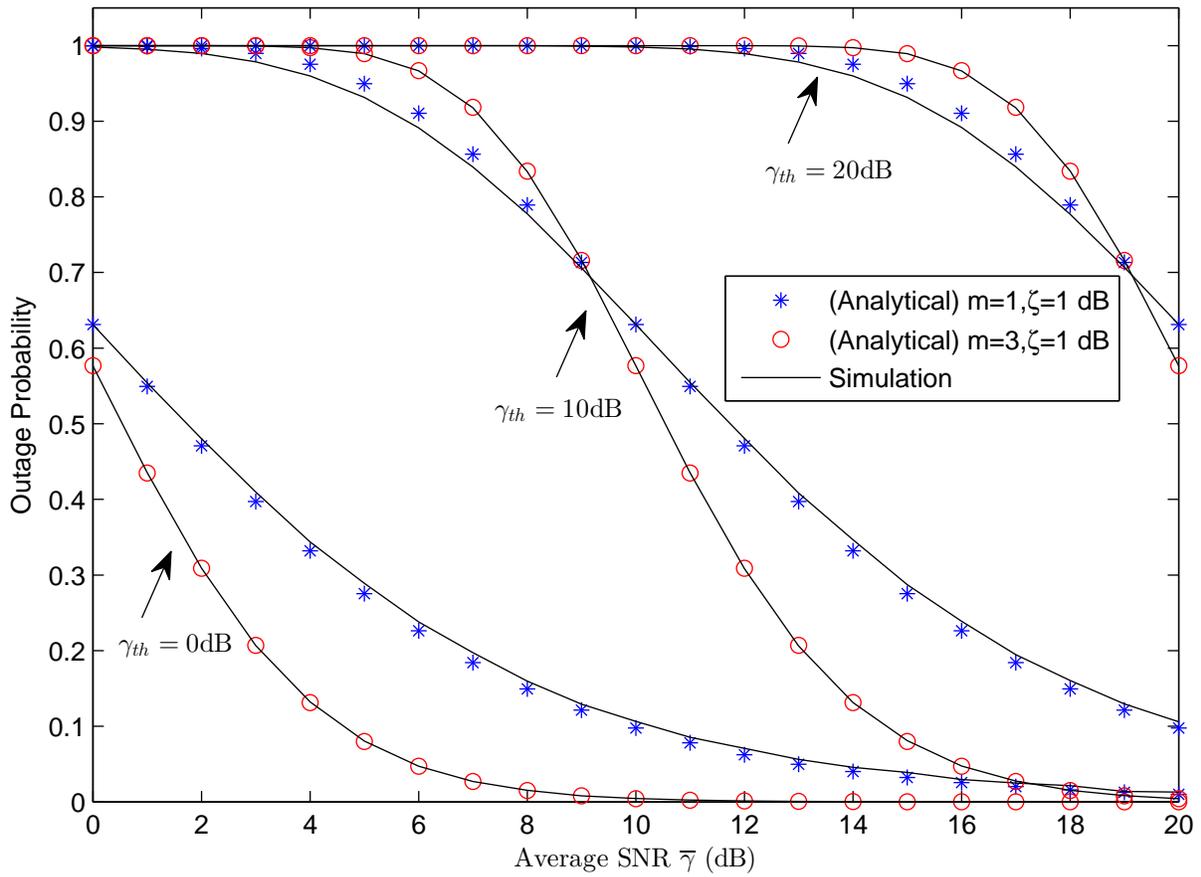

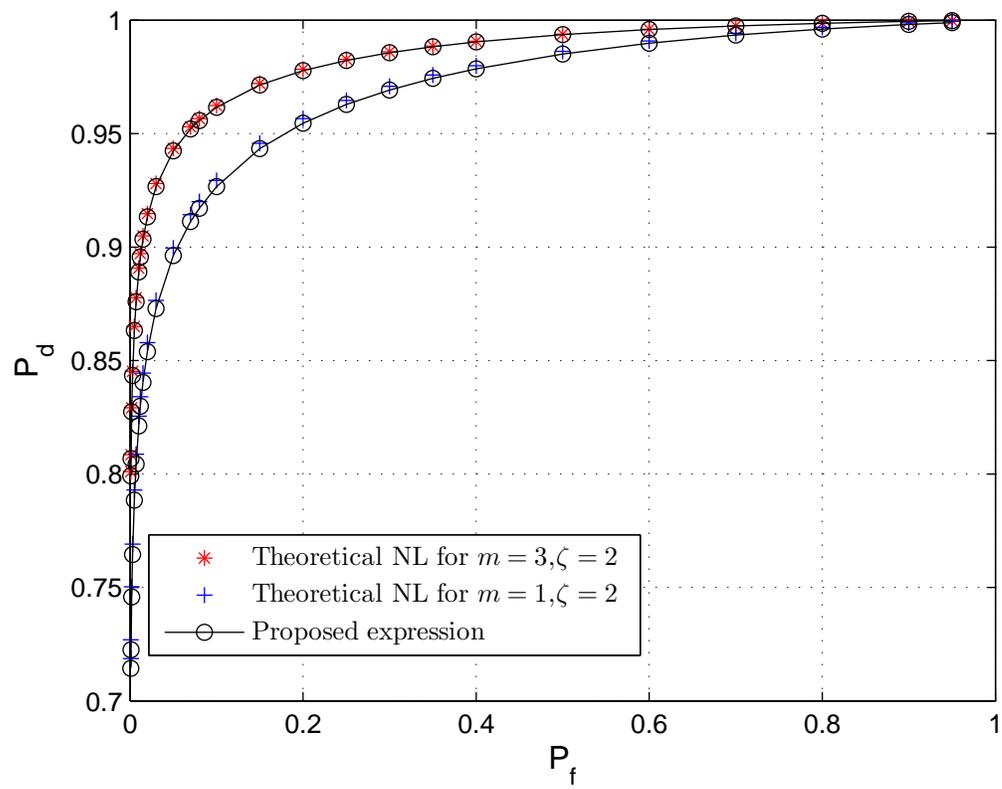